\documentclass[onecolumn,11pt]{article}
\usepackage{geometry}
\usepackage{amsmath}
\usepackage{amssymb}
\usepackage{color}
\usepackage{graphicx}
\usepackage{soul}
\usepackage{authblk}
\usepackage{siunitx}

\title{3D Printed Actuators: Reversibility, Relaxation and Ratcheting}

\author[1,2,4]{Song-Chuan Zhao \thanks{songchuan.zhao@outlook.com}}
\author[2]{Mariska Maas}
\author[2]{Kaspar Jansen}
\author[1,3]{Martin van Hecke}
\affil[1]{Huygens-Kamerling Onnes Laboratories, University of Leiden, Postbus 9504, 2300 RA, The Netherlands}
\affil[2]{Department of Design Engineering, Delft University of Technology, Landbergstraat 15, 2628 CE, Delft, The Netherlands}
\affil[3]{AMOLF, Science Park 104, 1098 XG Amsterdam, The Netherlands}
\affil[4]{Institute for Multiscale Simulation, Friedrich-Alexander-Universit\"{a}t, Caustra\ss e 3, 91058 Erlangen,Germany}

\date{}

\begin{document}

\maketitle

\begin{abstract}
Additive manufacturing strives to combine
any combination of materials into three dimensional functional structures and devices, ultimately
opening up the possibility of 3D printed machines. It remains difficult to actuate such devices,
thus limiting the scope of 3D printed machines to passive devices or necessitating the incorporation of external actuators
that are manufactured differently.
Here we explore 3D printed hybrid thermoplast/conducter bilayers,
that can be actuated by differential heating caused by
externally controllable currents flowing through their conducting faces.
We uncover the functionality of such actuators and show that they allow to 3D print, in one pass,
simple flexible robotic structures that propel forward under step-wise applied voltages.
Moreover, exploiting the thermoplasticity of the non-conducting plastic parts
at elevated temperatures, we show how strong driving leads to irreversible deformations - a form of 4D printing -
which also enlarges the range of linear response of the actuators.
Finally, we show how to leverage such thermoplastic relaxations to accumulate plastic deformations and
obtain very large deformations by alternatively driving both layers of a bilayer; we call this ratcheting.
Our strategy is scalable and widely applicable, and opens up a new approach to
reversible actuation and irreversible 4D printing of arbitrary structures and machines.
\end{abstract}

\section{Introduction}

4D printing allows the targeted temporal evolution of the shape, property and functionality
of a 3D printed structure in response to external stimuli \cite{Momeni2017,Kuang2018}. In particular
it enables the printing of flat structures, which upon actuation due to changes in the environment
morph into three dimensional objects, thus enabling self-folding of origami and self-assembly~\cite{Tibbits2014,Ge2013,Manen2018}.
In essence, this shape morphing is implemented by combining two or more materials
with significantly different material properties in the 3D printing process. The simplest example is a bilayer strip consisting of materials which differ in their ability to expand when exposed to external stimuli. Such a strip will bend or warp when exposed to changes in temperature~\cite{Ge2013, Wu2016}, moisture~\cite{Tibbits2014,Raviv2014}, light~\cite{Mu2015}
or pH~\cite{Nadgorny2016}. A well know natural example of this differential expansion mechanism is the opening and closing of a pine cone due to changes in humidity~\cite{SydneyGladman2016}.

Depending on the application in mind, the
shape morphing may be required to be {\em reversible}, for example when repeatedly actuating a structure,
or {\em irreversible}, for example when creating complex shapes from initially flat structures.
The differential expansion mechanism is, in principle, reversible, so that after removal of the stimulus, the structure should retake its original shape. In practice, repeatability is not always achieved, for example hydrogels used for moisture or temperature actuation suffering from degradation~\cite{Raviv2014}. One intriguing approach to
irreversible shape morphing was proposed by Ge {\em et al.}, based on bilayer structures with embedded shape memory polymer fibers ~\cite{Ge2013}. After printing,  a preconditioning step, consisting of heating, stretching and cooling,  pre-stretches the fibers and prepares them to actuate as soon as the temperature rises sufficiently.
Properly positioned end stops allow a better control of the end shape~\cite{Mao2015}. The procedure however requires manual pre-stretching before activation. Recently, integration of the pre-stretching step
into the 3D printing process has been demonstrated 
by controlled photopolymerization~\cite{Ding2017}.
In most cases, the materials are soft and fragile and cannot be used in load bearing structures. Moreover,
such shape morphing strategies relies on the diffusion of heat, moisture, or chemical signals. This make actuation
slow for macroscopic objects. In addition, such environmental triggers act on the entire object, making
actuation of individual parts a challenge; even though one can preprogram a given sequential activation by embedding soft materials with different critical thresholds (for example different glass transition temperatures) \cite{Manen2018}, full control of the actuation sequence can not be achieved by a globally diffusing signal.

To overcome these hurdles we present here a new actuator design which does not rely on  environmental triggers but instead allows for precise, spatiotemporal control of targeted regions in a 3D printed object.
The actuators consist of bilayers of thermoplastic material and conducting material, and
are activated by heating due to electrical currents which conveniently allows for control by external electronics.
We show that these actuators can act reversibly and return to their initial state, which we leverage to 3D print a self-propelling robotic ``turtle''.
Moreover, our actuators can  be driven towards an irreversible regime, allowing large deflections and externally triggered, permanent shape morphing. Finally, by inducing permanent deformations in both sides of a bilayer, we can achieve very large deformations that can be reversed by repeated irreversible deformations - we refer to this as {\em ratcheting}.

\section{Phenomenology}

\begin{figure*}
\centering
\includegraphics[width=17.8cm]{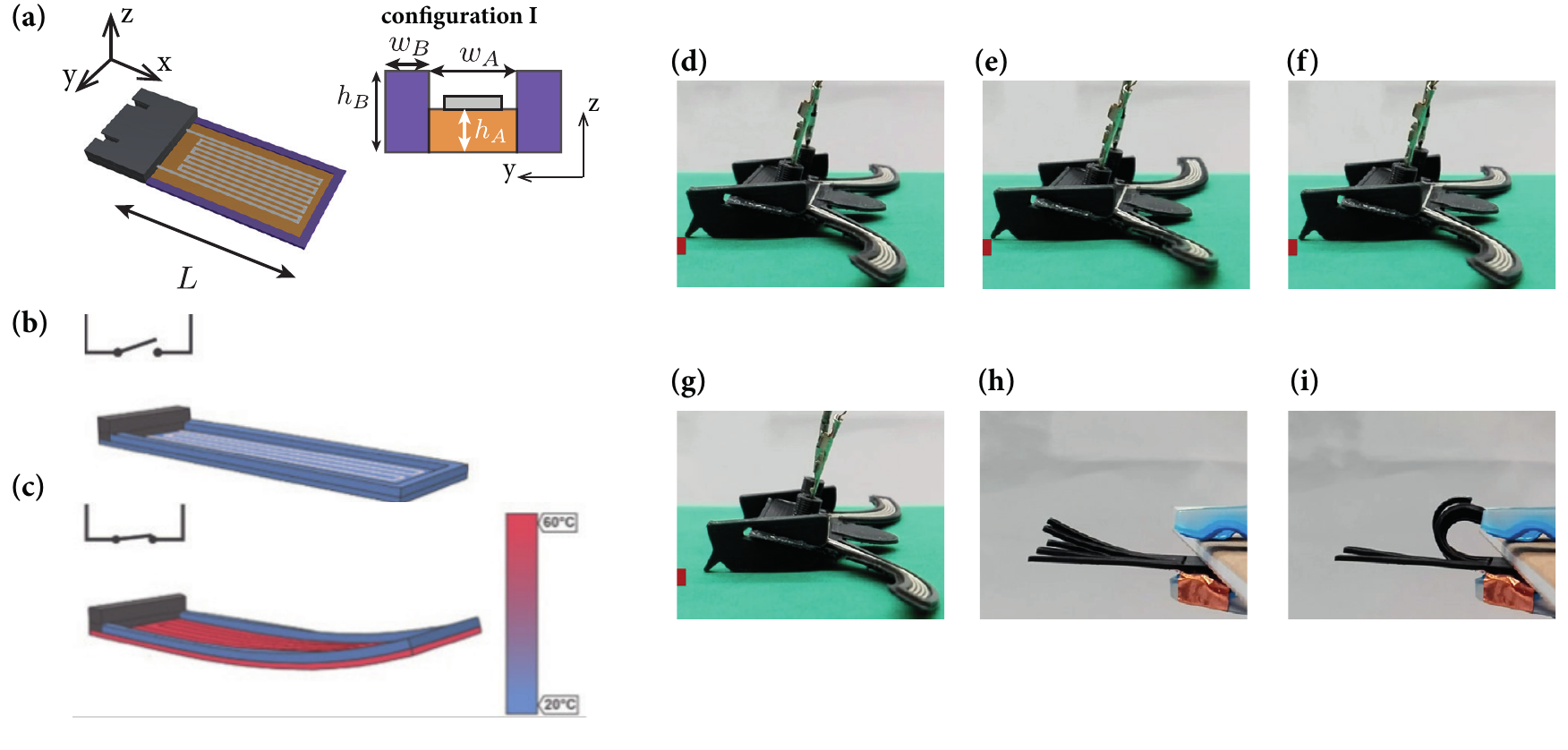}
\caption{\label{f.1} (a) Geometry of the base-rim structure and associated parameters; (b) Initial flat shape with cold base and cold rim; (c) Actuated curved shape with hot base and cold rim; (d) 3D printed turtle in the initial position with downward curved fins; (e) turtle with fins in actuation, upward position; (f) turtle at the end of the first actuation cycle; (g) turtle after 10 heating pulses; (h) Side view of a set of linear actuators arranged as fingers. (i) Strongly curved, plastically deformed fingers after strong actuation.}
\end{figure*}

We first demonstrate both reversible and irreversible actuation of 3D printed bilayer-based structures.
Our hybrid actuators consist of an insulating thermoplastic (PLA) and a conducting
ink (silver nano-particles gel)  and are produced on one-pass by a Voxel-8 3D FDM printer, which allows a resolution of {0.25 mm in the xy plane and of 0.15 mm in the z direction}. {The Voxel-8 is equipped with two extruder heads filled with thermoplastic material and the conductive ink respectively, allowing to simultaneously print PLA and the conductive ink to obtain the hybrid structures considered in this paper, without the need of manual interaction or post-processing. A demo video of the printing process can be found in the supplementary materials.} In configuration I, the body of this actuator consists of a base and a rim, with a heating circuit printed on top of the base (Figure~\ref{f.1}a-b). The geometry is specified by the widths $w_{A,B}$, thicknesses $h_{A,B}$ and length $L$; unless noted otherwise, we fix $h_A=0.4 $ mm, $h_B=0.8$ mm, $w_A=15$ mm and $w_B=1.5$ mm. We occasionally use configuration II, where the heating circuit is printed at the bottom of the base. A current send through the circuit predominantly heats the base, causing it to expand and leading to bending of the actuator (Figure~\ref{f.1}c). The function of the rim is to shift the neutral plane upwards, and exploratory tests indeed showed that a strip without a rim bends downwards when heated, whereas a strip with a rim bends upwards (see S.I.). In addition, later in the paper we explore the base-rim configuration to create a bilayer structure in which the base and rim can be heated independently. In the Supporting Information the choice of geometry and configurations is discussed.

We have 3D printed a turtle-shape robot whose front fins contain the base-rim structure described above. Upon sending a moderate current through the actuators, the base heats up, and the actuator bends and lowers the robot. When the current is interrupted, the fins cool down and the combination of lowering and raising the robot
results in a small forward motion (Figure~\ref{f.1}d-g and SI video). Ten current pulses drive the tiny robot forward by approximately 3.5 mm, without irreversible changes in the shape of the fins: this demonstrates the reversible actuation of our bilayers.
We have also printed a structure with five actuators, arranged as fingers, where
the heater is printed below the base and the rim is on top (Figure~\ref{f.1}h). We
have investigated the deformations that occur when the actuators are subjected to significantly larger currents, which cause heating above the glass transition temperature of the PLA. We observe very large and irreversible, bending deformations of the fingers that are heated (Figure~\ref{f.1}i). Together, these two examples demonstrate the feasibility and applicability of both reversible and irreversible heating strategies for a single type of actuator.

\section{Reversible and irreversible actuation}

To understand the role of reversible and irreversible deformations in our actuator, we have performed experiments where we drive freshly printed samples through a series of heating-cooling cycles. As
the body of the sample is made of a thermoplastic that expands upon heating but also
relaxes and flows for temperatures approaching its glass transition, complex behavior can be expected to arise.
Our experiments will reveal such relaxation, and moreover find evidence for a
`recoiling'-like relaxation that occurs whenever the sample is first heated above the glass transition.

\begin{figure*}
\centering
\includegraphics[width=17.8cm]{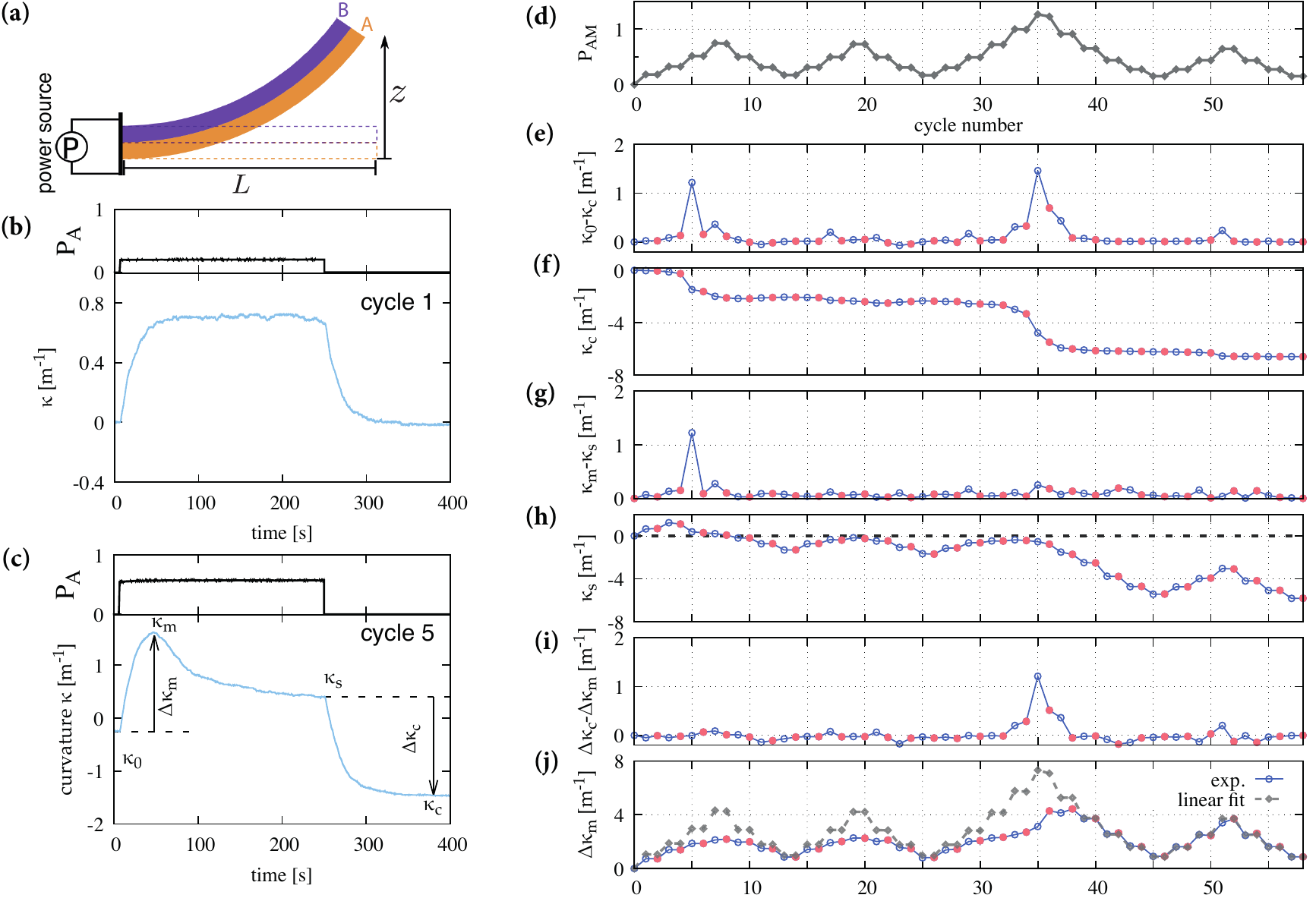}
\caption{\label{f.exp} (a) Bilayer analogy with a base (A) and rim (B). (b) A heating cycle leading to reversible bending for $P_{AM} \ll 1$.
(c) A heating cycle leading to irreversible bending cycle for $P_{AM} \approx 0.6$, with characteristic parameters  $\kappa_0$, $ \kappa_m$,  $ \kappa_s$, $ \kappa_c$,
$ \Delta\kappa_m$ and $\Delta\kappa_c$ indicated.
(d) A freshly printed sample undergoing a series of heating cycles governed by $P_{AM}$. For each $P_{AM}$ two equal heating cycles are implemented, with  the first cycle plotted in open blue circles, and the second in solid orange circles. Cycle 1 and 5 are shown in panel (b) and (c) respectively.
(e) The time trace of  $\kappa_0 - \kappa_c$ signifies irreversible behavior around cycle 5 and cycle 35.
(f-j) Time traces of $\kappa_c$, $\kappa_m-\kappa_s$, $\kappa_s$,  $\Delta \kappa_c - \Delta\kappa_m$ and $\Delta\kappa_m$ reveal the different nature of the irreversible behavior around cycle 5 and cycle 35 (see text).
In panel (j), we compare the data (blue and red circles) to a proportional response to $P_{AM}$, with the proportionality constant obtained from a fit to cycles 40 and later.
}
\end{figure*}

We subject a single actuator (Figure~\ref{f.1}a) through a series of heating-cooling cycles. While the details of the design of the printed sample
are important, the main physics can be interpreted best via a bilayer analogy (see S.I.), where layer $A$ models the heated base and layer $B$ the cold rim (Figure~\ref{f.exp}a). We probe the deflection at the tip of the actuator, $z$,  with a laser displacement sensor (MICRO-EPSILON ILD1302) and by taking side-view video images. We found that the bending curvature is uniform in most experiments, and report the approximate curvature
$\kappa := 2z/L^2$. All experiments are performed at ambient temperature {which is between 22\si{\celsius} and 24\si{\celsius}}.
Using infrared imaging, we have verified that both the temperature difference and temperature of the hot layer
is linear in the applied power $P_{exp}$. We have measured the critical power $P_g$ where layer A reaches the glass temperature $T_g=70^\circ$ C, and report in all plots the rescaled power $P_A:=P_{exp}/P_g$.

We monitor the curvature $\kappa(t)$ during heating cycles consisting of a 250 s heating period with power $P_A=P_{AM}$, followed by a 150 s cooling period with $P_A=0$. These periods are significantly larger than the thermal equilibration time which is of the order of tens of seconds (Figure~\ref{f.exp}b-c). For low heating power the phenomenology is simple: during heating the actuator bends upwards and  $\kappa(t)$ smoothly reaches a plateau, and during cooling, the actuator returns to the flat position where $\kappa \approx 0$ (Figure~\ref{f.exp}b). However, for larger heating power non-monotonic behavior is observed, and after cooling the actuator curves downwards in its cold state ($\kappa<0$), which signifies plastic deformations (Figure~\ref{f.exp}c). As we will show below, such irreversible deformations can be used to optimize the actuators response, and can be leveraged to yield new functional behavior that we refer to as ratcheting.

\subsection{Stress relaxation}

To understand relaxation at high heating power, we have performed experiments where we drive freshly printed samples through a series of heating-cooling cycles. Our experimental protocol consists
of 58 heating cycles, where we ramp the heating power of each cycle, $P_{AM}$, up and down four times
(Figure~\ref{f.exp}d). For each value of $P_{AM}$, we perform two subsequent thermal cycles. In the first two ramps the power peaks at $P_{AM}={0.7}$,
significantly below the critical power, whereas in the third ramp the heating power peaks at $P_{AM}={1.2}$, i.e., above the critical power; the fourth ramp is identical to the first two.
For all 58 cycles, we measure the curvature at the beginning of each cycle, $\kappa_0$, the maximum deflection during heating, $\kappa_m$, the final deflection during heating, $\kappa_s$, and the final deflection after cooling, $\kappa_c$ (Figure~\ref{f.exp}c). We define the maximal actuation
$ \Delta\kappa_m:\kappa_m-\kappa_0$ and the amount of bending during cooling $\Delta\kappa_c:= \kappa_s-\kappa_c$. For reversible actuation as in panel (b),
$\kappa_c =\kappa_0$, $\kappa_m=\kappa_s$, and $ \Delta\kappa_m =\Delta\kappa_c$.
We detect irreversible, plastic deformations of the actuator in particular in  cycle
5 and in cycle 35 (Figure~\ref{f.exp}e-i). As we will discuss in the sections below, the nature of these relaxation events is significantly different.

We  first focus on cycle 5. Here we observe a significant peak in $\kappa_m-\kappa_s$ but not in $ \Delta\kappa_c- \Delta\kappa_m$, and the deflection during the heating cycle, $\kappa_s$, relaxes to nearly zero. This strongly suggests that the relaxation in cycle 5 is driven by the relaxation of the bending induced stresses by plastic deformations of the hot layer A. Consistently with this, we observe a negative post-heating curvature $\kappa_c$.
To understand why the relaxation in cycle 5 is so much stronger than in cycle 1-4, we note that the rate of stress relaxation of thermoplastics increases rapidly for temperatures approaching the glass transition. Our data indicates that for cycle 5, where the heating power is larger than in previous cycles, all heating induced stresses relax during the heating period. In the subsequent cooling period, the base layer contracts, resulting
in a negative $\kappa_c$ (downward curvature), and $\Delta\kappa_c$ is of the same magnitude as the curvature change during heating. The system is now stressed in the cold state, but here the stresses cannot relax.
This picture is consistent with the behavior in cycle 6, which has the same heating power as cycle 5. During this cycle, the actuator
moves upwards towards an essentially flat and presumably stress-free state, and no stress relaxation and associated peak in $\kappa_c-\kappa_0$ is observed.
Consistent with this picture, there is some additional relaxation in cycle 7 but not in 8 nor in all subsequent cycles.

\subsection{Recoiling}

The relaxation that occurs in cycle 35 when $P\approx 1$ is of a different nature than in cycle 5; even though  $\kappa_c$ gets significantly more negative, $\kappa_m-\kappa_s$ does not show a significant peak, but rather $\Delta \kappa_c - \Delta\kappa_m$ peaks. Hence, the cooling leads to a much larger
deformation than the heating. Therefore, the second plastic deformation peak must have a different origin. We propose that the deformations in cycle 35 are due to the relaxation of frozen-in orientation which is released when the temperature first exceeds the glass transition (i.e. at cycle 35). The concept of flow induced molecular orientation is well known in the field of polymer processing~\cite{Osswald2003} and also occurs when the polymer melt is forced through the 3D print nozzle and subsequently solidifies. Heating such a sample with frozen-in orientation has two effects: first of all there is an entropic retraction force of the oriented molecules which it is proportional to $k_BT$ and thus increases with temperature~\cite{Treloar1975}. At low temperatures the polymer is in its glassy state and the retraction force is too small to cause any visible deformation. When approaching the glass transition the matrix stiffness rapidly drops by two to three orders of magnitude such that the oriented molecules (as well as sample layer A) now can contract, resulting in large downward curvature (Figure~\ref{f.exp}f). This effect is irreversible.

Consistent with this irreversible change in the microscopic structure,  we observe that the maximum deflection as function of $P$ after a heat pulse above the critical temperature
changes from a nonlinear to a linear function (see S.I. for elastic finite element simulations), significantly improving the sensitivity and functionality of the actuator - see Figure~\ref{f.exp}j. In a fresh sample the thermal expansion is counteracted by the contraction force of the oriented molecules. This contraction force is larger at higher temperatures (see cycles 5-10 and 13-17) but no longer present after the orientation is released (cycles 46-57).

In conclusion, our data shows that we can distinguish two  distinct mechanisms that drive irreversible deformations: First,
whenever the sample reaches a higher temperature than before this causes relaxation of the stresses and curvature: the hottest samples are straight and unstressed. Second, whenever the sample reaches for temperatures comparable to $T_g$ for the first time, there is an irreversible change in the microstructure of the plastic --- this change only happens once. {This one-time effect has been used for irreversible shape morphing~\cite{Zhang2015}. In the second part of this manuscript we will discuss a strategy of programmable irreversible deformations, \emph{ratcheting}, where recoiling is released by heating and does not take place any more.}

\section{Modeling}

We now introduce a simple model that qualitatively captures the reversible and irreversible
deformations as seen in, e.g., cycle 5 of the set of experiments discussed above; we do not attempt to
model the recoiling.

\begin{figure*}[h]
\centering
\includegraphics[width=12.5cm]{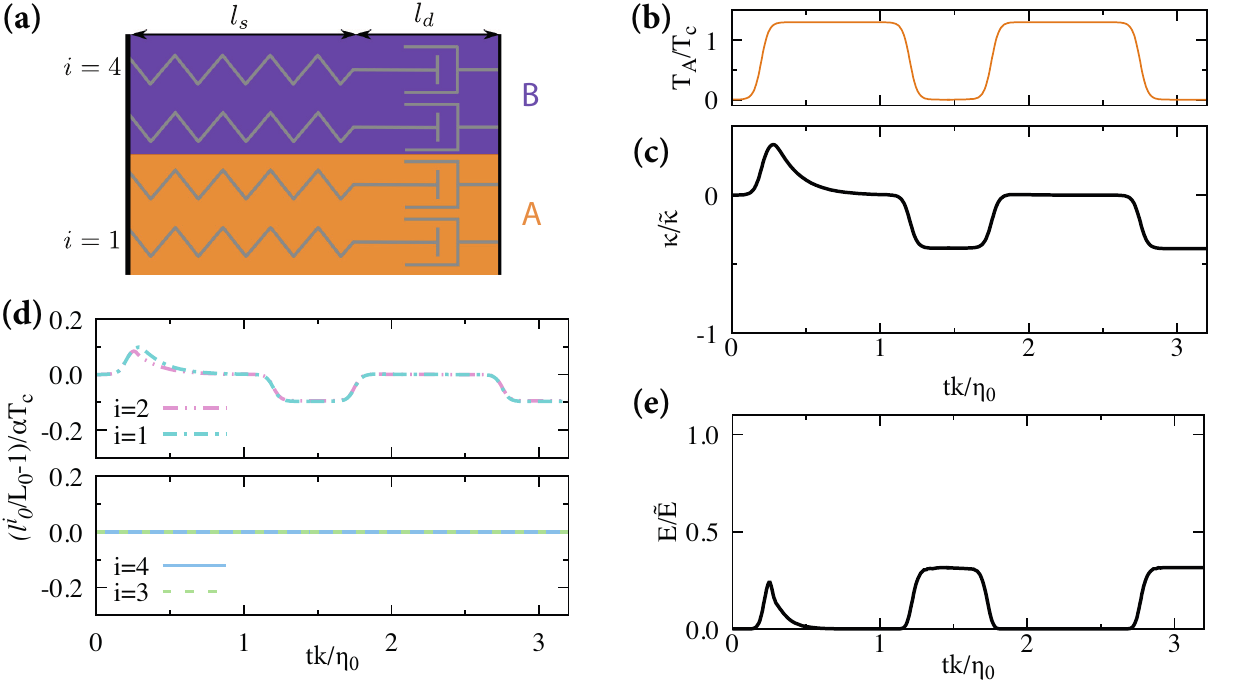}
\caption{\label{f.model} (a) The quadruple-spring model; each visco-elastic spring consist of a temperature dependent
elastic spring with length $l_s$ and temperature dependent viscous element with length $l_v$ - see text for details.
(b) Two subsequent heating pulses. (c) The non-dimensional curvature $\kappa/\bar{\kappa}$ shows an overshoot,  relaxation, and plastic deformations during the first cycle, and reversible actuation in subsequent cycles.
(d) Evolution of the (virtual) unstressed spring lengths $l^i_0/L_0$. (e) Evolution of the stored elastic energy normalized by $\tilde{E}=\frac{1}{2}k(\alpha T_cL_0)^2$.
}
\end{figure*}

We model the bilayer actuator with four Maxwell-type visco-elastic compound springs, two for each layer
(Figure~\ref{f.model}a).
We assume these springs to be equidistant in the transverse $z$-direction with spacing $d$.
Each compound spring has a length $l^i$ (where $i=1,\dots 4$), and
consist of an elastic spring with spring constant $k$  and length $l_s^i$,
in series with a viscous element with a temperature dependent time scale $\eta(T)$ and length $l_v^i$.
The rest length of the elastic springs depends on temperature as
$l^i_{s0}(T_i)=(1+\alpha T^i)L_{0}$, where
$\alpha$ is the thermal expansion coefficient, $T^i$ the temperature of spring $i$, and
 $L_{0}$ the rest length at $T=0$.
We model the temperature dependence of the timescale of the viscous element by expanding the
Williams-Landel-Ferry Equation,
$\eta(T)/\eta_0=10^{{ \frac{c_1(T-T_c)}{c_2+(T-T_c)}}}$ near $T_c$, yielding
\begin{equation}
\eta(T) = \eta_0 10^{{A(1-T/T_c)}}   ~,
\end{equation}
and fix $A=7$ to ensure that forces completely relax within the heating time; the precise choice of the temperature dependence is not critical.

The temperature dependence of both the springs rest length and timescale allow for actuation and plastic relaxation respectively .
Taken together, each compound spring evolves according to a simple ODE:
\begin{equation}
\eta(T) ~ d{l_v^i}/d_t  = k\left(l_{s0}^i(T)-l_s^i\right)~. \\
\label{eq.inner_force}
\end{equation}
{We assume that during heating, the two springs in the heated layer have equal, positive temperature $T_A(t)$,
while the two springs in the non-heated layer stay at $T=0$}. We
assume a linear gradient in the extensions of the compound springs: $l^i-l^{i-1}=\beta d$, leading to bending of the actuator, and we define the curvature $\kappa:=
\beta/\langle l^i\rangle$.
For a given temperature evolution $T^i(t)$, we solve the evolution of $\beta$ by balancing {forces and moments}.
{To monitor the plastic deformations, we below report $l_0^i$, the (virtual) length of the compound spring under zero force, which combines thermal expansion and accumulated visco-plastic relaxations of the spring.}
We non-dimensionalize our data by the characteristic scales $d$ (length), $\eta_0/k$ (time),  $T_c$ (temperature) and $kd^2$ (energy), and introduce a characteristic curvature scale $\bar{\kappa}$ as $(\alpha T_c)/d$.

We report the curvature and length of the compound springs as function of time for
%
%
two subsequent  heating pulses (Figure~\ref{f.model}b-e).
For  lower heating powers, the system simply actuate reversibly (not shown). Here we focus on $T_A=1.3T_c$, and this high power ensures that the system has significant viscous relaxation. We observe that during the first heating cycle, the curvature displays a clear overshot, and then
relaxation towards the neutral position - very similar to what we saw in the experiments (Figure~\ref{f.model}c). Clearly, this relaxation stems from the viscous relaxation of the compound springs during the heating cycle (Figure~\ref{f.model}d);
we also observe that the stored elastic energy decays to zero during the heating phase of the first cycle (Figure~\ref{f.model}e).
Subsequent cooling leads to a stressed, oppositely curved configuration, while additional heating cycles with the same peak temperature drive the system back to a flat, relaxed state in the hot phase{, where $l^3_0=l^4_0$} (Figure~\ref{f.model}c-e). {Hence, irreversible changes in the rest-length of the springs at high temperatures drive
relaxation of the stresses during the hot phase of the cycle. In turn, these lead to a cold state that is stressed and non-flat, and a phenomenology that is
consistent with the experimental observations of relaxation (c.f. Figure~\ref{f.exp} for cycles 5 and 6).}

\section{Ratcheting}

\begin{figure*}
\centering
\includegraphics[width=17.8cm]{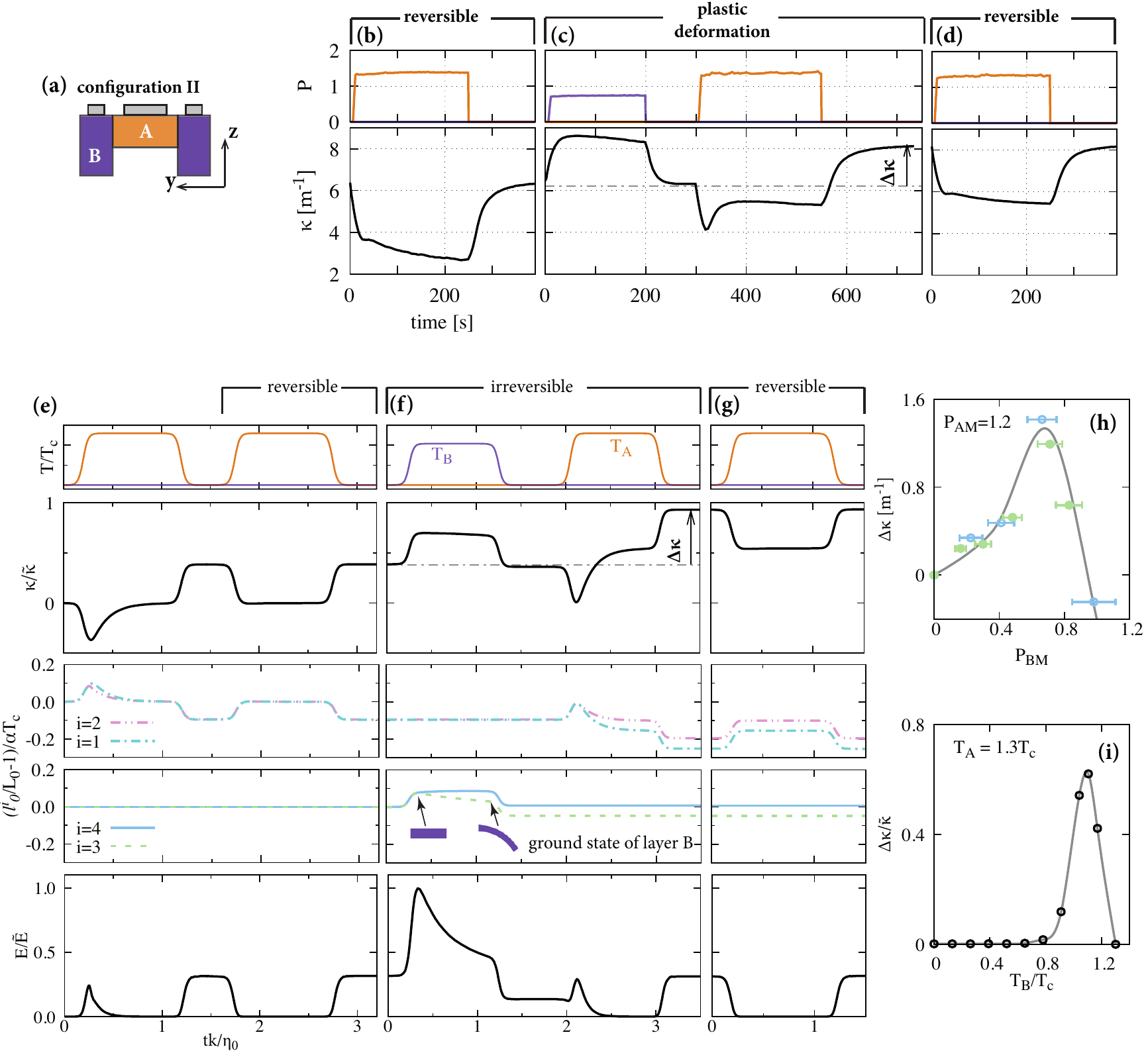}
\caption{\label{f.ratch_1cycle}
(a) A cross-section of our actuator with heating layers at both layer A and B. (b)
Reversible actuation by heating layer A - previous heating cycles of layer A have already caused plastic deformations.
(c) Subsequent heating of layer B and layer A lead to additional plastic deformation of the cold state characterized by
$\Delta k$. Notice that $P_{BM}<P_{AM}$.
(d) Subsequent heating of layer A leads to reversible behavior, but with  a larger deformation in the cold state than in panel (b).
(e-g) Result of model simulations of a similar heating  protocol: (e) Reversible actuation by heating layer A, where previous heating cycles have caused plastic deformations. (f) Subsequent heating of layer B and layer A lead to additional plastic deformation of the cold state. The elastic energy, {normalized by $\tilde{E}=\frac{1}{2}k(\alpha T_cL_0)^2$, shows that the system is not fully relaxed. The `unstressed' spring lengths show that the ground state of layer B is changed from flat to curved.}
(g) Subsequent heating of layer A leads to reversible behavior, but with  a larger deformation in the cold state than in panel (e).
(h) The experimentally determined plastic deformation, $\Delta\kappa$ peaks as function of $P_{BM}$ for fixed $P_{AM}=1.2$. (i) Similarly, in the model the plastic deformation $\Delta\kappa/\bar{\kappa}$ peaks as function of $T_B$ for fixed $T_A=1.2 T_c$.
}
\end{figure*}

The maximal reversible deflection of the actuators that we have considered so far is limited by the amount of heating that we can use before plastic deformations set in. Here we show that we can leverage a complex exchange of deformations and stresses between two layers to obtain large and reversible deformations by a process that we refer to as {\em ratcheting}. For this, we aim at alternatively heating layer A and B, and thus print a layer of conducting ink on the rim, so that we independently can heat either layer (Figure~\ref{f.ratch_1cycle}a).
Note that we now employ a type II configuration, where the conducting ink is printed at the surface of the actuator;
experimentally, we found that this configuration is less sensitive to developing transversal curvatures which for large deformations can {warp and ultimately} destroy the samples (See S.I.). In addition, notice that the heating layer is now on top, so that actuation {of layer $A$ leads to downward motion.}

We now consider a sample that has received multiple large heating pulses at layer A, so that
it has undergone plastic deformations, bends upwards in the cold state, and has a reversible response (Figure~\ref{f.ratch_1cycle}b). {Note that the recoiling effect has been removed by heating the layer above $T_g$.}
In this configuration, layer B is under tensile stresses, but these cannot relax as long as
layer B remains cold - what happens when we send a heating pulse to layer B? {We focus here on the scenario where $T_{BM}$ is large enough to cause some  plastic deformations during heating, but not full relaxation of the stresses.}
Our data shows that {such} heating {of} layer B has two effects. First, as expected, the actuator curves even more during heating, and during cooling it relaxes back to a state with a curvature that is close to its preheating curvature.
However, we observe that a subsequent
heating pulse in layer A leads to additional plastic deformations and even larger curvature of the actuator in the cold state (Figure~\ref{f.ratch_1cycle}c). The different response of the actuator before and after the heating pulse to layer B
implies that  the internal state of the actuator must have changed during the heating of layer B.
Additional heating of layer A do not lead to further plastic deformations (Figure~\ref{f.ratch_1cycle}d).

To understand the mechanism that drives the complex response to alternately heating both layers in our actuators, we have performed a similar heating strategy in our model where we explicitly can monitor the {plastic deformations,} stress relaxations and elastic energies. As in the experiment, we prepare the model into a state that has
undergone large heating pulses in layer A (Figure~\ref{f.ratch_1cycle}e; note that with respect to Figure~\ref{f.model}, we now reverse layer A and B to be consistent with the current set of experiments).
When we apply heating to layer B, this leads
to a deformation that appears nearly reversible as far as the curvature is concerned, similar to the experiments (Figure~\ref{f.ratch_1cycle}f). {However, a close inspection of the evolution of the (unstressed) spring lengths $l_0^i$ and stored
elastic energy, reveals that the rest lengths $l_0^3$ and $l_0^4$ become unequal, and that the elastic energy rises sharply before slowly relaxing (Figure~\ref{f.ratch_1cycle}f). Clearly, after the heating pulse of layer $B$ has ended, the elastic energy has not fully relaxed,  and the unstressed state of layer $B$ (``the ground state'') is changed from flat to curved. Similar to the experiments, subsequent heating cycles
of layer A cause a significant, irreversible increase in the post-heating curvature in the first cycle, and reversible behavior in subsequent cycles (Figure~\ref{f.ratch_1cycle}g).

Taken together, our experiments and model illustrate the core idea of ratcheting, where alternately heating layer A  and B {with different heating powers, we can leverage partial relaxation of the actuator to  accumulate} large deformations. To optimize the amplitude of these deformations, we
have experimentally studies the amount of ratcheting $\Delta\kappa$ as function of $P_{BM}$ for fixed $P_{AM}$.
We can anticipate that no ratcheting can occur for
very low $P_{BM}$, while for
very  large $P_{BM}$, additional flow in layer B negates the ratcheting. Consistent with this, we
observe a peak in $\Delta\kappa$ for $P_{BM} \approx 0.7$ (Figure~\ref{f.ratch_1cycle}h). Similarly, in the model, the amount of ratcheting
$\Delta \kappa/\bar{\kappa}$ peaks for intermediate heating of the B layer (Figure~\ref{f.ratch_1cycle}i).
Hence we conclude that alternately heating layer A and B incurs irreversible deformations of our actuator if $P_{AM}$ and $P_{BM}$ are chosen appropriately.

\begin{figure*}
\centering
\includegraphics[width=17.8cm]{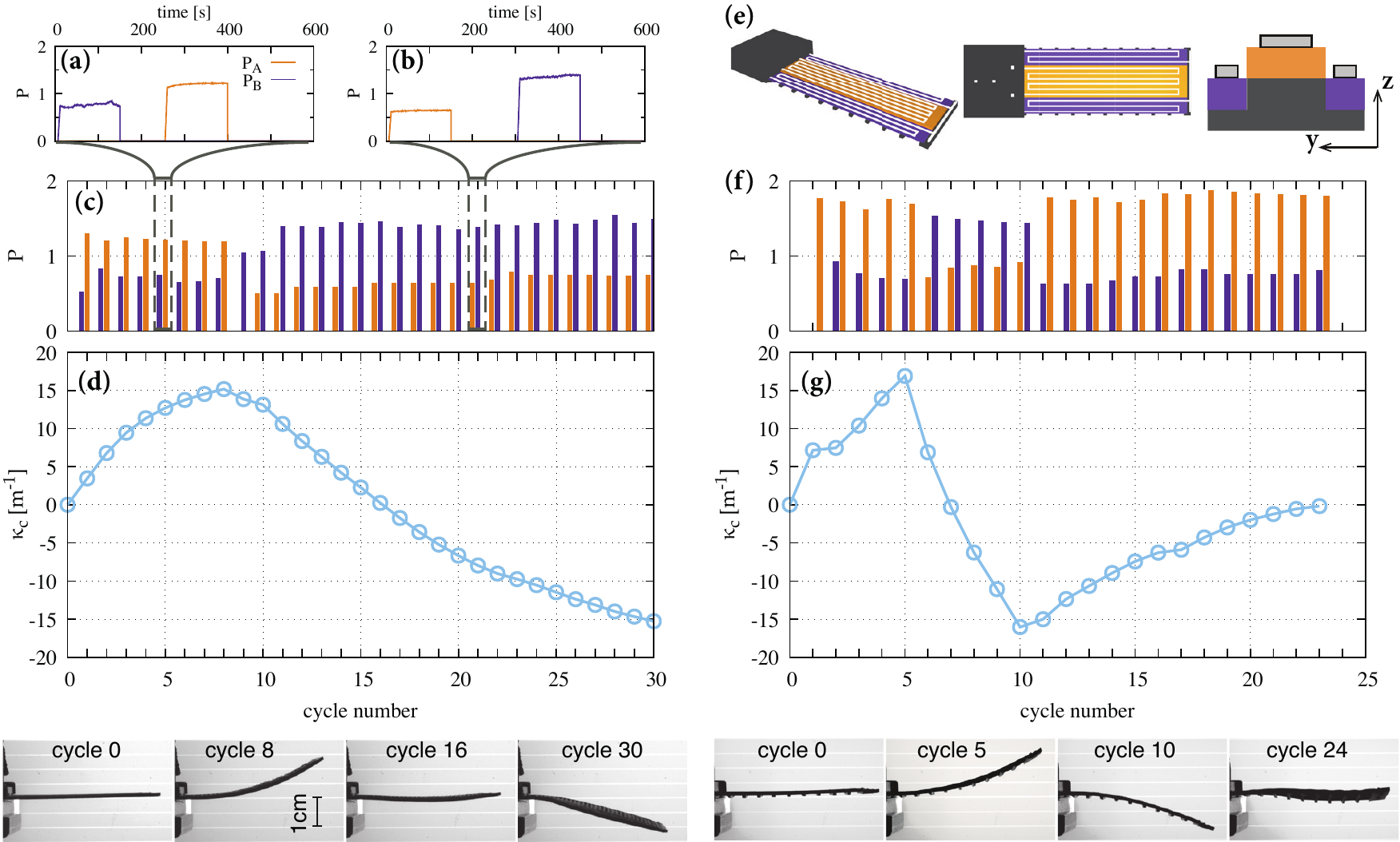}
\caption{\label{f.ratchet_exp}(a-c) Complex driving protocol, where alternately $P_{AM}$ and $P_{BM}$ dominate.
(d) Multiple driving pulses lead to large cumulative deformations (see cycle 8) that can be completely reversed (cycle 16) and even be inverted (cycle 30). (e) Different actuator design which is less sensitive to lateral warping. (f) Actuation protocol --- the duration of the heating pulses is 150 s similar to (a) and (b). (g) Rapid, large and reversible deformations.}
\end{figure*}

We now show that multiple ratcheting cycles can lead to large cumulative deformations, and that by controlling both $P_{AM}$ and $P_{BM}$ we can control the direction of ratcheting, thus allowing large but ultimately reversible deformations.
We have performed experiments for a series of ratcheting heating pulses (Figure~\ref{f.ratchet_exp}). We find that repeated alternating heating cycles lead to a steady increase of the curvature, far beyond what can be achieved in a single cycle. Moreover, by inverting the relative strength of $P_{AM}$ and $P_{BM}$, we can change the direction of motion, go back to an essentially flat state, and also reach strongly negatively curved samples (Figure~\ref{f.ratchet_exp}d). The reader may refer to the Supporting Information for finite element simulations and additional details.
Finally, additional experiments  with an improved actuator design (Figure~\ref{f.ratchet_exp}e)  show that
it is possible to return to the flat state also after having experienced both large positive and negative curvatures (Figure~\ref{f.ratchet_exp}f-g). Hence, heating protocols where multiple areas are actuated in sequence open the door
to very large and easily controllable  deformations of 3D printed hybrid materials. 

\section{Conclusion and Outlook}

In this paper we have introduced a new design for 3D printed actuators that deform upon electrical heating. Depending on the driving strength, both reversible and irreversible deformations can be realized. We demonstrated reversible actuation by a one-pass 3D printed flexible robot turtle that propels forward under repeated actuation. We demonstrated  irreversible,  large deformations leading to strongly curved strips, and note that this allows a novel strategy of electrically activated 4D printing. Finally, we demonstrate ratcheting:  by alternately heating both layers at appropriately chosen temperatures,  irreversible deformations  that remain fixed in the absence of driving
can be accumulated, but these deformations can also be erased and reversed with subsequent heating cycles. This shows that electrical heating of thermoplastics forms a flexible platform for reversible actuation, 4D printing, and a combination of the two.

We show that a model based on spring-dashpots captures the complexity of our actuators,  
reproduces the `ratcheting' experiments, and thus uncovers the basic mechanisms.
While the thermoplastic material used in the current work is polylactide, 
the principles of our actuation strategies are expected to apply to thermoplastic materials in general. Our work thus opens up a new avenue for shape morphing and actuation of 3D printed structures. 

\section{Acknowledgment}
We acknowledge funding from the Netherlands Organization for Scientific Research a VICI grant No. NWO-680-47-609.

\section{Conflict of Interest}
The authors declare no conflict of interest.

\end{document}